\documentclass[twocolumn, times, twocolappendix]{aastex631}

\usepackage{color}
\usepackage{xfrac}

\newcommand {\valrmin} {2.3}
\newcommand {\valrmax} {13.2}
\newcommand {\Nsample} {30}

\newcommand {\valgamma} {3.398}
\newcommand {\valegamma} {0.094}

\newcommand {\valbeta} {-0.72}
\newcommand {\valepbeta} {0.62}
\newcommand {\valembeta} {1.07}

\newcommand {\valalpha} {0.21}

\newcommand {\valf} {1.00}
\newcommand {\valepf} {0.16}
\newcommand {\valemf} {0.15}

\newcommand {\valMrmax} {2.66}
\newcommand {\valepMrmax} {0.42}
\newcommand {\valemMrmax} {0.36}

\newcommand {\valMvir} {1.80}
\newcommand {\valepMvir} {1.05}
\newcommand {\valemMvir} {0.54}

\newcommand {\valrvir} {148}
\newcommand {\valeprvir} {24}
\newcommand {\valemrvir} {16}

\newcommand {\valc} {12.6}
\newcommand {\valepc} {2.2}
\newcommand {\valemc} {2.3}

\newcommand {\valvrmax} {92.9}
\newcommand {\valepvrmax} {6.2}
\newcommand {\valemvrmax} {5.7}

\newcommand {\kms} {km\,s$^{-1}$}
\newcommand {\masyr} {mas\,yr$^{-1}$}
\newcommand {\Msun} {M$_\odot$}

\newcommand {\rmin} {r_\mathrm{min}}
\newcommand {\rmax} {r_\mathrm{max}}
\newcommand {\rvir} {r_\mathrm{virial}}
\newcommand {\Mvir} {M_\mathrm{virial}}
\newcommand {\vcirc} {v_\mathrm{circ}}
\newcommand {\Mrmax} {M \left( < \rmax \right)}
\newcommand {\Mvalrmax} {M \left( < \valrmax \, \mathrm{kpc} \right)}
\newcommand {\Mmodelrmax} {M_\mathrm{model} \left( < \rmax \right)}

\newcommand {\Mtme} {M_\mathrm{TME}}
\newcommand {\Mtrue} {M_\mathrm{true}}
\newcommand {\Pvc} {P_{vc}}
\newcommand {\Pmass} {P_\mathrm{mass}}
\newcommand {\Phalo} {P_\mathrm{halo}}
\newcommand {\Mlmc} {M_\mathrm{LMC}}

\newcommand {\vt} {\overline{v_\tau}}

\newcommand {\st} {\sigma_\tau}

\newcommand {\vrr} {\overline{v_r^2}}
\newcommand {\vaa} {\overline{v_\phi^2}}
\newcommand {\vpp} {\overline{v_\theta^2}}
\newcommand {\vtt} {\overline{v_\tau^2}}

\graphicspath{{./}{Figures/}}

\shorttitle{The Mass of the Large Magellanic Cloud}
\shortauthors{Watkins et al.}

\begin{document}

\title{The Mass of the Large Magellanic Cloud from the Three-Dimensional Kinematics of its Globular Clusters}

\correspondingauthor{Laura Watkins}
\email{lwatkins@stsci.edu}

\author[0000-0002-1343-134X]{Laura L. Watkins}
\affiliation{AURA for the European Space Agency (ESA), ESA Office, Space Telescope Science Institute, 3700 San Martin Drive, Baltimore, MD 21218, USA}

\author[0000-0001-7827-7825]{Roeland P. van der Marel}
\affiliation{Space Telescope Science Institute, 3700 San Martin Drive, Baltimore, MD 21218, USA}
\affiliation{Center for Astrophysical Sciences, The William H. Miller III Department of Physics \& Astronomy, Johns Hopkins University, Baltimore, MD 21218, USA}

\author[0000-0001-8354-7279]{Paul Bennet}
\affiliation{Space Telescope Science Institute, 3700 San Martin Drive, Baltimore, MD 21218, USA}

\begin{abstract}
We estimate the mass of the Large Magellanic Cloud (LMC) using the kinematics of 30 LMC globular clusters (GCs). We combine proper motions (PMs) measured with \textit{HST}, \textit{Gaia}, or a combination of the two, from a recent study by \citet{Bennet2022} with literature line-of-sight velocities (LOSVs) to give 3 components of motion. With these, we derive a 3D velocity dispersion anisotropy $\beta = \valbeta ^{+\valepbeta} _{-\valembeta}$, consistent with the GCs forming a flattened system with significant azimuthal motion. We then apply a tracer mass estimator and measure an enclosed mass $\Mvalrmax = \valMrmax ^{+\valepMrmax} _{-\valemMrmax} \times 10^{10}$~\Msun. This is broadly consistent with results from previous studies of the LOSVs of GCs and other luminous tracers. Assuming a cosmologically-constrained NFW distribution for the dark matter, this implies a virial mass $\Mvir = \valMvir ^{+\valepMvir} _{-\valemMvir} \times 10^{11}$~\Msun . Despite being an extrapolation by almost an order of magnitude in radius, this result is consistent with published estimates from other methods that are directly sensitive to the LMC's total mass. Our results support the conclusion that the LMC is approximately 17$^{+10}_{-6}$\% of the Milky Way's mass, making it a significant contributor to the Local Group (LG) potential.
\end{abstract}

\keywords{Large Magellanic Cloud (903), Galaxy masses (607), Galaxy mass distribution (606), Galaxy dynamics (591), Galaxy kinematics (602), Globular star clusters (656)}

\section{Introduction}
\label{section:introduction}

The Large Magellanic Cloud (LMC) is the largest of the dwarf galaxies orbiting the Milky Way (MW). So large, in fact, that the moniker of dwarf galaxy is a bit of a misnomer. It sits in a very interesting niche within the Local Group (LG), being both sufficiently different in mass than the MW to be an interesting comparison and sufficiently massive to be a major player in the MW's recent history and present state. 

Mass is a fundamental quantity of any stellar system. The total amount of mass and its distribution govern how the system behaves and how the stars inside it move. The mass both determines and is determined by its formation and past evolution, and dictates its likely future.

Mass is also a reference point. We can study nearby galaxies in much greater detail than we can those more distant, and so we use our knowledge of the nearby universe to help interpret observations of more distant systems. Mass is a metric commonly used to compare and connect systems near and far.

The MW is our first, best benchmark, with a mass $\sim 1.1 \times 10^{12}$~\Msun \citep[e.g.][]{Patel2018, Deason2019, Watkins2019, Cautun2020, Sawala2023}. As our home and, by definition, the closest galaxy to us, it is the best studied and most convenient reference point. The mass of nearby neighbor Andromeda (M31) is of the same order of magnitude as the MW, albeit possibly slightly larger \citep[e.g.][]{Watkins2010, Fardal2013, VillaneuvaDomingo2021, Patel2023}. This is useful in providing perspective of variation in galaxies of similar mass, but does not help for understanding galaxies of different mass. Most of the MW and M31 dwarf galaxies are several orders of magnitude lower in mass \citep{Collins2014}, providing a very different mass reference point, but there is a big mass jump between them and their hosts.

There are a few local galaxies that sit in between the two (local) extremes: they are, in approximate mass order, M33 \citep{Corbelli2014}, the LMC \citep{Erkal2021, Shipp2021}, M32 \citep{DSouza2018}, and the Small Magellanic Cloud (SMC) \citep{Stanimirovic2004, DeLeo2023}. M33 and M32 are at approximately the same distance as M31, which limits the detail with which we can study them, both in terms of data accuracy and data type. The LMC and SMC are much closer, a likely-bound pair of galaxies on their first infall into the MW \citep[e.g.][]{Kallivayalil2013}, with the LMC being approximately an order of magnitude less massive than the MW, and the SMC an order of magnitude less massive again. This makes both the LMC and SMC very useful mass benchmarks.

The LMC is especially interesting as it is massive enough to have a significant and observable effect on the MW. \citet{Laporte2018} showed that the warp in the MW's disk could be entirely explained as a response to the infalling LMC. \citet{Garavito2019} showed that the LMC should leave a clear wake as it moves through the MW halo, seen both spatially and in kinematics, and \citet{Petersen2020} showed that the LMC can move the MW disk barycentre, which can in turn produce reflex motion in the MW halo. \citet{Erkal2020} used equilibrium models to show that failure to account for the impact of the LMC on the MW can bias MW mass estimates considerably. The response of the MW to the LMC's infall has now been observed in multiple studies, including \citet{Belokurov2019}, \citet{Petersen2021}, \citet{Conroy2021}, and \citet{Erkal2021}. But the extent of the LMC's influence, and whether it is indeed solely responsible for some of these features depends upon its mass.

The LMC can also influence the tracks of stellar streams in the MW halo. \citet{Law2010} used the Sagittarius (Sgr) stream to constrain the shape of the MW halo; they found halo properties that were unexpected in a cold dark matter (DM) paradigm, and concluded that this could be due to the influence of the LMC. \citet{VeroCiro2013} made a similar study and concluded that the dynamical perturbations induced by the LMC on Sgr's orbit must be accounted for in modelling the stream tracks. Later, \citet{Koposov2019} and \citet{Shipp2019} identified such perturbations using Gaia PMs for the Orphan Stream and for a handful of MW streams respectively. \citet{Gomez2015} showed that the MW responds to the gravitational pull of the LMC, and that both the LMC itself and the MW response can affect the Sgr stream. \citet{Erkal2018} also found that Tucana III has passed very close to the LMC, and that the LMC can have a significant effect on its orbit, with the degree of the effect determined by the LMC's mass.

Like most galaxies, the LMC is believed to be DM dominated, with visible matter making up only a small fraction of its total mass. We turn to dynamics to estimate the mass, inferring both the total amount of mass and its distribution by observing either the relative motion of the LMC and other nearby objects, or by studying the motions of objects inside it.

To date, most studies have done the former. \citet{Kallivayalil2013} studied the orbit of the LMC and SMC pair around the MW; \citet{Penarrubia2016} applied the timing argument to the MW-M31-LMC system; \citet{Erkal2019}, \citet{Vasiliev2021} and \citet{Shipp2021} used the effect of the LMC on MW stellar streams; and \citet{Laporte2018} and \citet{Erkal2021} measured the mass via the MW response to the LMC.

Here, we do the latter. The motions of the tracers depend on and can, thus, be used to estimate the mass interior to the most distant tracer. The limitation to such a method is that visible tracers tend to be much more limited in extent than the DM, so do not directly provide estimates of the total mass. For example, \citet{vdMarel2014} used LMC stars as tracers to estimate the mass of the LMC within 8.7~kpc. However, estimates of the mass within any given radius are still extremely useful for understanding the internal structure of the LMC, and, as we will see, with some judicious assumptions, we can make a reasonable extrapolation of the total mass as well.

The most accurate estimates require knowledge of full 6D phase space information for the tracers, that is, positions on the plane of the sky, distances, line-of-sight velocities (LOSVs), and proper motions (PMs). Recently, \citet{Bennet2022} measured PMs\footnote{In part using \textsc{GaiaHub}, a public tool for combining \textit{HST} and \textit{Gaia} data to measure PMs \citep{delPino2022}.} for a set of LMC globular clusters (GCs). They supplemented this with literature LOSVs, distances, and some additional PM measurements, to compile a catalog of 32 LMC GCs with 6D phase space information. This catalog is ideal for estimating the mass of the LMC, and we do so here.

In \autoref{section:dataprocessing}, we describe the \citet{Bennet2022} catalog in more detail and how we extracted from it the information needed for this study. In \autoref{section:results}, we describe the tracer mass estimator method and then apply it to the LMC to estimate the mass within the most distant tracer, and to extrapolate the total LMC mass. In \autoref{section:discussion}, we discuss our results in the context of previous studies. In \autoref{section:conclusions}, we summarise our results.

\section{Cluster Sample}
\label{section:dataprocessing}

We begin with the catalog of 32 GCs in the vicinity of the LMC from \citet{Bennet2022}. This provides 6D phase space information --- positions on the sky, distances, PMs, and LOSVs --- for all clusters.

\begin{deluxetable*}{cccc}
\tablecaption{Adopted LMC Properties and Sources \label{table:lmcprops}}
\tablehead{
    \colhead{Property} &
    \colhead{Symbol} &
    \colhead{Value} &
    \colhead{Source}
}
\startdata
Centre & ($\alpha_0$, $\delta_0$) & (81.28$^\circ$, -69.78$^\circ$) & \citet{vdMarel2001} \\
Distance & $D$ & 50.1 $\pm$ 2.5~kpc & \citet{vdMarel2002} \\
LOSV & $v_{LOS}$ & 262.2 $\pm$ 3.4 \kms & \citet{McConnachie2012} \\
PM & $\mu_W$ & -1.910 $\pm$ 0.020 \masyr & \citet{Kallivayalil2013} \\
& $\mu_N$ & 0.229 $\pm$ 0.047 \masyr & \\
Disk inclination & $i$ & $37.4^\circ \pm 6.2^\circ$ & \citet{vdMarel2001} \\
Position angle of the line of nodes & $\Theta$ & $-129^\circ \pm 8.3^\circ$ & \citet{vdMarel2001} \\
\enddata
\end{deluxetable*}

From these, we wish to calculate LMC-centric separations $r$ and velocities $v$, for which we need the properties of the LMC. The LMC is a disk, so in addition to its position and velocity, we need the orientation of the disk as well. The adopted properties, along with their sources, are provided in \autoref{table:lmcprops}.

Given the large size of the LMC on the sky, small angle approximations do not hold, so the relative positions and velocities of the clusters are calculated using full spherical trigonometry. We follow the prescriptions and equations laid out in \citet{vdMarel2001} and \citet{vdMarel2002}, which give positions and velocities of the clusters in a cartesian coordinate system centred on the LMC, where the $x$-axis lies in the disk along the line of nodes, the $y$-axis lies in the disk perpendicular to the line of nodes, and the $z$-axis is perpendicular to the disk. Then $r$ is the distance from the centre of the LMC, and $v$ is the total velocity relative to the LMC. From these cartesian coordinates, we also calculate positions and velocities in both spherical and cylindrical coordinate systems centred on the LMC.

To propagate uncertainties on these values, we assume that the LMC and cluster positions on the sky are fixed, but sample all other observable LMC and cluster properties (distances, LOSVs, PMs, disk angles) assuming the uncertainties are Gaussian. We draw 10,000 samples for each cluster, then calculate the LMC-centric values for each sample. The resulting distributions are often not Gaussian, so we adopt the medians as the best estimates, and the 15.9 and 84.1 percentiles\footnote{For a perfectly Gaussian distribution, these percentiles would represent the 1-$\sigma$ confidence interval.} as the uncertainties.

\begin{figure}
    \centering
    \includegraphics[width=\linewidth]{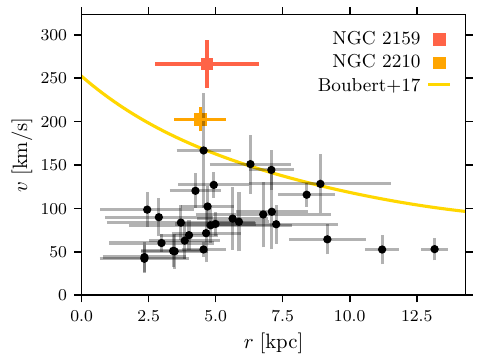}
    \caption{LMC-centric positions and distances for 32 clusters in the vicinity of the LMC. Likely-unbound clusters NGC 2159 and NGC 2210 are highlighted in red and orange respectively. The yellow line shows the escape velocity curve from \citet{Boubert2017} that we used to distinguish likely-bound and likely-unbound clusters.}
    \label{figure:data}
\end{figure}

The distribution of LMC-centric distances $r$ and velocities $v$ for all 32 clusters is shown in \autoref{figure:data}. Two clusters are highlighted: NGC\,2159 in red and NGC\,2210 in orange. We can see that these two clusters have very high velocities given their distance from the centre of the LMC, higher than the escape velocity of the LMC at their distance \citep[yellow line,][]{Boubert2017}. This indicates that they are likely not bound to the LMC and do not trace the mass of the LMC. Their kinematics are also very different from the rest of the sample, further suggesting they are not part of the LMC cluster population. For both of these reasons, we remove these clusters from our sample.

This leaves us with a sample of $\Nsample$ clusters. The minimum distance $\rmin$ in the sample is $\valrmin$~kpc, and the maximum radius $\rmax$ (within which we will measure the mass) is $\valrmax$~kpc.

\section{LMC Mass}
\label{section:results}

\citet{Watkins2010} described a set of simple tracer mass estimators (TMEs), that is, they estimate the mass of an object given a set of tracers within the radius of the most distant tracer $\rmax$. The estimators assume that, over the region of interest $\rmin$ to $\rmax$, the underlying potential can be described as a power-law with index $\alpha$, the tracers are distributed following a power-law with index $\gamma$, and the tracer population has a constant velocity anisotropy $\beta$.

Despite their simplicity, these estimators have proved to be remarkably effective, even in messy, complex systems. \citet{Watkins2010} used them to estimate the masses of both the MW and M31 using their dwarf galaxy populations; \citet{Annibali2018} estimated the mass of LMC-analogue NGC\,4449 using its globular clusters; and \citet{Sohn2018} and \citet{Watkins2019} estimated the mass of the MW using \textit{HST} and \textit{Gaia} PMs of its globular clusters.

The estimators work with different types of distance and velocity measures, depending on the kind of data available for a given system. Here, we have full 6D phase space information, so we use the estimator that works with 3D distances $r$ and velocities $v$,
\begin{equation}
    \Mrmax_\mathrm{TME} = \frac{\alpha+\gamma-2\beta}{G \left( 3-2\beta \right)} \rmax^{1-\alpha} \left\langle v^2 r^\alpha \right\rangle .
	\label{eqn:TME}
\end{equation}
First, we will estimate $\alpha$, $\beta$, and $\gamma$. Then we will apply the TME to the sample.

\subsection{Tracer Density}
\label{section:tracerdensity}

We start by fitting a power-law function to the tracer density profile. This is the density of all objects in the population from which the tracers were drawn, not just those for which velocities have been measured. So we take our original sample of 32 clusters and augment it with a further 5 clusters for which sky positions and distances are available but which lack a LOSV measurement, a PM measurement, or both. For the 5 additional clusters, we draw samples of their LMC-centric distances, as described in \autoref{section:dataprocessing}. This gives us a set of 37 clusters, each with 10,000 sampled LMC-centric distances $r$.

We use the full samples for each cluster, and calculate the cumulative number profile, normalized so that the total number of clusters is 37. We then perform a least-squares fit, assuming a power-law density over the region, to recover the power-law index $\gamma$ of the density profile.

\begin{figure}
    \centering
    \includegraphics[width=\linewidth]{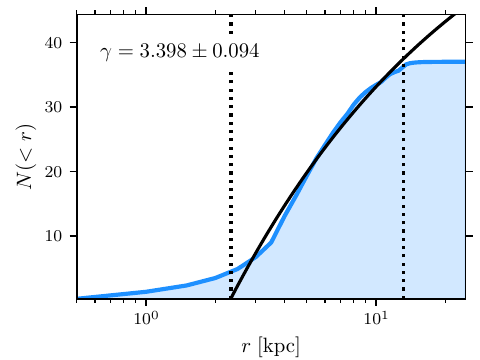}
    \caption{Cumulative number profile of LMC GCs. Dotted lines mark the inner ($\rmin$) and outer ($\rmax$) radii of the main sample at $\valrmin$ and $\valrmax$ kpc. The black line is a power law density fit over the region of interest. The power-law index of the fit, $\gamma$, is shown in the top left corner. The quoted error bar is that for the fit only, and does not include all possible sources of uncertainty.}
    \label{figure:tracerdensity}
\end{figure}

\autoref{figure:tracerdensity} shows the normalized cumulative number profile of all 370,000 samples in blue. The vertical dotted lines delineate the region of interest at $\rmin$ and $\rmax$. The solid black line shows the cumulative profile for an underlying power-law density with index $\gamma$. We find a best-fitting value for $\gamma = \valgamma \pm \valegamma$, which we will assume is fixed, and use for the analysis.

\subsection{Anisotropy}
\label{section:anisotropy}

To calculate the anisotropy, we start with the 10,000 samples of \Nsample\ clusters that we drew in \autoref{section:dataprocessing}. From these, we wish to estimate the second velocity moments of the cluster population relative to the LMC in spherical coordinates: radial $\vrr$, azimuthal $\vaa$, and polar $\vpp$.

Typically, we would proceed by calculating the mean and standard deviation of the 10,000 samples for each velocity component for each cluster, and then do a maximum-likelihood fit, assuming everything is nicely Gaussian, to estimate the first and second velocity moments.

The problem here is that everything is not nicely Gaussian. Some clusters do have velocity components whose distributions are very well approximated by a Gaussian, but others do not at all. This is impactful in two ways: first, that when we calculate a mean and standard deviation of the sample, these values are not particularly representative of the distribution; and second, that when we assume the distribution is Gaussian with the mean and standard deviation we calculated, we are further compounding the problem.

This is inconvenient because likelihood functions are typically convolutions of two Gaussians, and Gaussians are straightforward to convolve. Introducing non-Gaussian distributions can be highly non-trivial. However, convolution is distributive, so fitting the distributions as a sum of $N$ Gaussians will still give us simple likelihood functions.

Now we wish to determine what value of $N$ is sufficient to represent the shapes of the distributions. To each of the 3 velocity components for each of the 30 clusters, we fit a single Gaussian $G_1$, a linear sum of two Gaussians $G_2$, and a linear sum of three Gaussians $G_3$. That is,
\begin{equation}
    G_N = \sum^N_k f_k G \left( m_k, w_k \right) ,
    \label{equation:gaussians}
\end{equation}
for Gaussian functions $G$ with mean $m_k$ and width $w_k$. The fractions $f_k$ must sum to 1, so in practice, we fit $f_1$ to $f_{N-1}$ and calculate $f_N$ such that
\begin{equation}
    f_N = 1 - \sum^{N-1}_k f_k .
\end{equation}
Upon visual inspection, we found that some distributions were indeed fit well by a single Gaussian, but many were not. The fits were improved using two Gaussians, though some distributions were still not fit well. Using three Gaussians seems to be good enough in all cases. For consistency, we use a three-Gaussian fit for all distributions.

\begin{figure*}
    \centering
    \includegraphics[width=\linewidth]{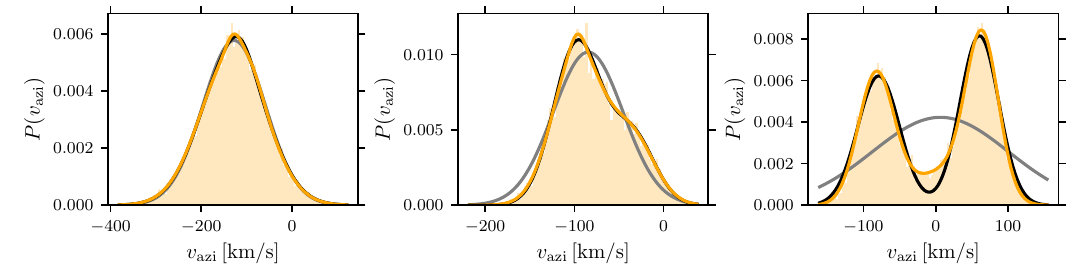}
    \caption{Distribution of azimuthal velocities for 3 clusters in our sample highlighting the (in)adequacy of different Gaussian fits. The orange histograms show the distributions of the 10,000 sample draws. One-Gaussian fits are shown as grey lines, two-Gaussian fits are shown as black lines, and three-Gaussian fits are shown as orange lines. The shapes of the distributions are discussed in the text. \textbf{Left:} NGC\,1652. The grey and black lines are hardly visible, indicating that all 3 fits are very similar. This is a case where a one-Gaussian fit would be sufficient. \textbf{Middle:} NGC\,1831. The black line is hardly visible under the orange line, indicating that these fits are very similar, but the grey line is distinct and is a clearly poor fit to the data. This is a case where a two-Gaussian fit would be sufficient. \textbf{Right:} NGC\,1916. The three fits are clearly distinct. The two-Gaussian fit provides a much better fit than the one-Gaussian fit, but doesn't fit the central dip well, which the three-Gaussian fit does. This is a case where the three-Gaussian fit is needed.}
    \label{figure:veldbns}
\end{figure*}

\autoref{figure:veldbns} shows azimuthal velocity distributions for three clusters as examples for each of the three cases. The histograms show the distributions of the 10,000 samples, and the grey, black and orange lines show the one-, two- and three-Gaussian fits; sometimes the lines are hard to see where the fits are very similar and the lines overlap. The leftmost panel shows NGC\,1652 and is a case where a one-Gaussian fit would be adequate. The middle panel shows NGC\,1831 and is a case where the two-Gaussian fit would be adequate. The rightmost panel shows NGC\,1916 and is a case where the three-Gaussian fit is needed.

The reason for the double-peaked distributions, such as we see for NGC\,1916, is due to geometry. For some clusters, when we Monte Carlo sample both the cluster properties and the LMC properties, some sampled points put the cluster on one side of the LMC centre and the remaining points put the cluster on the opposite side of the LMC centre. Moving the cluster's position relative to the centre changes the sign of the velocity.

Now we are ready to estimate the first and second moments of the cluster population. Our model assumes that the population has mean velocity $\vt$ and dispersion $\st$ for in each coordinate $\tau = \{ r, \theta, \phi\}$. Then the likelihood of the data given the model is
\begin{equation}
    \mathcal{L} = \prod_\tau \prod_j \left[ \sum^3_k f_{\tau jk} G \left( m_{\tau jk} - \vt, \sqrt{w_{\tau jk}^2 + \st^2} \right) \right] .
    \label{equation:likelihood}
\end{equation}
for velocity components $\tau$, clusters $j$, and Gaussian components $k$ (which have been previously fit and are not fitted here).

We assume flat priors for the mean velocities in each component. We require that the dispersions are positive, but otherwise use a flat prior for the dispersions. The posterior is then the product of the likelihood and the priors.

To explore the parameter space of the means and dispersions, and to find the region for which the posterior is maximized, we use the affine-invariant Markov Chain Monte Carlo (MCMC) package \textsc{emcee} \citep{ForemanMackey2013}; we use 200 walkers for 1500 steps, after which time the runs appear to have fully converged -- that is, the parameter space appears to be well and evenly sampled. We draw 10\,000 points from the final posterior distribution for our final sample. For each parameter, we adopt the median as the best estimate and use the 15.9 and 84.1 percentiles for the uncertainties.

From the means $\vt$ and dispersions $\st$, we can then estimate the second velocity moments
\begin{equation}
    \vtt = \st^2 + \vt^2
\end{equation}
in each coordinate. Then finally, we are able to calculate the anisotropy
\begin{equation}
    \beta = 1 - \frac{\vaa + \vpp}{2 \vrr}
	\label{eqn:anisotropy}
\end{equation}
of the system.

\begin{figure}
    \centering
    \includegraphics[width=0.8\linewidth]{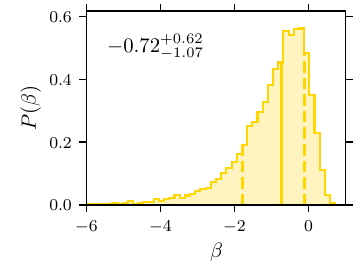}
    \caption{Distribution of anisotropy values. There is a long negative tail to the distribution due to the highly asymmetric nature of $\beta$ as defined, so we have truncated the figure to highlight the region with the majority (99.6\%) of the sample.}
    \label{figure:anisotropy}
\end{figure}

\autoref{figure:anisotropy} shows the resulting distribution of $\beta$ values. The anisotropy is negative, indicating that the tangential velocities are larger than the radial, as expected in a flattened, rotating, pressure-supported system and consistent with the findings in \citet{Bennet2022}.

This distribution is non-Gaussian, and so in what follows we will draw directly from this distribution to better sample this parameter. However, it is useful to offer a best estimate. We find $\beta = \valbeta^{+\valepbeta}_{-\valembeta}$, where the best estimate is taken as the median of the distribution, and the 15.9 and 84.1 percentiles are used for the uncertainties. The vertical solid and dashed lines show the best estimate and uncertainties in the figure.

\subsection{Potential}
\label{section:potential}

For the potential, we assume that the LMC is composed of a disk embedded in a massive halo. In principle, the underlying structure is irrelevant for our mass estimates, we need only the power-law slope of the combined potential. However, we would like to extrapolate our mass $\Mrmax$ out to estimate the virial mass $\Mvir$, and use measurements of the circular velocity in the disk to make sure our approach is consistent with observations, so we do need to take some care in considering the potential.

The region spanned by our cluster sample is fairly central, and is a region over which a disk would certainly be relevant. We adopt a double exponential disk model, using a disk mass of $2.5 \times 10^9$~\Msun \citep{Kim1998}, a disk scale length of 1.15~kpc \citep{Saha2010}, and a scale height that is $\sfrac{1}{3}$ the scale length \citep{Alves2004}. The disk properties are held fixed.

We assume a spherical NFW halo, defined by concentration $c$ and virial radius $\rvir$. If we knew $c$ and $\rvir$, then we would not need to do this work, as that would define the mass enclosed at all radii within $\rvir$. So we will sample a range of halos to account for this uncertainty.

We will describe the sample in \autoref{section:halos}. Here we simply describe how we fit a power-law to the potential for any given ($c$, $\rvir$) combination.

We need to calculate the potential as a function of radius over the region of interest. The halo is spherical, so this is straightforward. The disk is not spherical so, any time we need a disk potential, we draw an angle at random in $0 \le \sin \theta \le 1$ to account for non-sphericity. The angle above the plane is not a significant source of uncertainty on $\alpha$, but we do sample in angle nevertheless. We sum the disk and halo potentials to get the total potential, and fit to that a power law over the region of interest. Then $\alpha$ is the slope of the power-law fit.

\begin{figure}
    \centering
    \includegraphics[width=\linewidth]{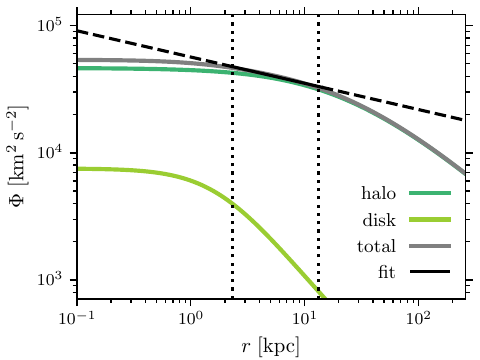}
    \caption{Potential profiles used in this study. The light green line shows the double exponential disk potential, which is fixed. The dark green line shows the potential of an NFW halo, the NFW parameters are allowed to vary in the study and a single example is shown here. The grey line shows the total potential, which is the sum of the disk and the halo. The vertical dotted lines mark the region of interest. The black line shows a power-law fit to the total potential over the region of interest; the line is solid in the region of interest and dashed outside of the region of interest.}
    \label{figure:potentials}
\end{figure}

\autoref{figure:potentials} shows an example of the adopted potentials and a power-law fit. The fixed disk potential is shown in light green, here plotted for an angle above the disk plane of 30$^\circ$. An example NFW halo potential is shown dark green for $c =$ 13 and $\rvir =$ 160~kpc. The sum of the disk and halo potentials is the shown in grey. The vertical dotted lines delineate the region of interest. The black line shows a power-law fit to the total potential, solid within the region of interest where the fits are performed, and dashed outside.

\subsection{Halo Sample}
\label{section:halos}

As mentioned in \autoref{section:potential}, we sample a range of NFW halos in ($c$, $\rvir$) space. Now we need to set the properties of the sample. We use a concentration range $6 \le c \le 20$ and a virial radius range $80 \le \rvir \le 240$~kpc, which encompasses the range of values that is plausible for a galaxy like the LMC (as we discuss further below). We choose to sample at random across this space; virial mass $\Mvir$ is determined solely by $\rvir$ and is independent of $c$, and we will achieve more robust statistics later by sampling $\Mvir$ more fully than using a regular grid. For robust statistics, we sample 50,000 halos across this space.

However, not all halos in this space are equally likely. We can weight the halos to account for prior knowledge.

To begin, we consider how consistent the circular velocity of the model is with observations. \citet{vdMarel2014} measured the circular velocity at 8.7~kpc to be $\vcirc$(8.7~kpc) = 91.7 $\pm$ 18.8~\kms. It would be helpful to have another estimate at a different radius. \citet{GaiaCollaboration2021} measured both rotation curves and rotation velocity dispersion profiles for a number of stellar populations in the LMC disk. Together, and with a suitable correction for asymmetric drift, these can be used to estimate the circular velocity at any radius covered by the study. We chose to estimate the circular velocity at 4~kpc, as this is still within the range of our clusters, but is at a little under half of the \citet{vdMarel2014} value, and provides a unique anchor.

We use the two youngest populations from \citet{GaiaCollaboration2021}; we choose the youngest as they should have the smallest asymmetric drift corrections, and we choose two populations to get an indication of the uncertainty on the circular velocity. For population Young2, $v_\phi$(4~kpc) = 88~km/s and $\sigma_\phi$(4~kpc) = 9~km/s. For population Young3, $v_\phi$(4~kpc) = 74~km/s and $\sigma_\phi$(4~kpc) = 17~km/s\footnote{We read these values from the plots in Figure B.5 of \citet{GaiaCollaboration2021}. There are three sources of uncertainty: those shown on the plots (negligible), human error in estimating from plots (modest), and scatter across different populations (considerable). The latter is by far the dominant source of uncertainty, and is the driver for why we sampled two populations here.}. We follow the formalism described in \citet{vdMarel2002}, to correct for asymmetric drift, using a correction factor $\kappa = 3.5$. This gives us circular velocities $v_\mathrm{circ} = 90$~km/s for Young 2 and $v_\mathrm{circ} = 80$~km/s for Young 3. Finally, we combine this information and adopt $v_\mathrm{circ}$(4~kpc) = 85 $\pm$ 5~km/s as our estimate of the circular velocity.

\begin{figure}
    \centering
    \includegraphics[width=\linewidth]{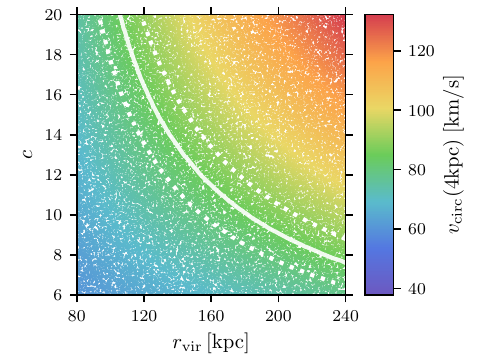}
    \includegraphics[width=\linewidth]{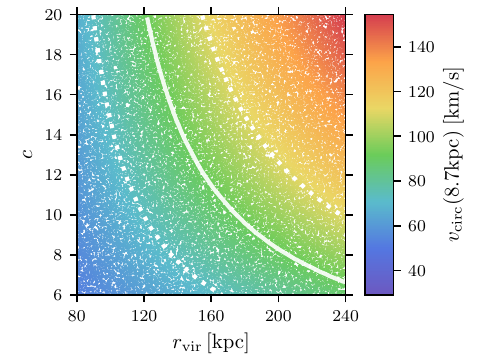}
    \caption{Circular velocities at 4~kpc (top) and 8.7~kpc (bottom) of sampled halos. The color indicates the circular velocity as indicated by the color bars to the right. For guidance, the solid white lines indicate the observed values and the dotted white lines indicate the uncertainties on the observed values.}
    \label{figure:halos_vcircs}
\end{figure}

\autoref{figure:halos_vcircs} shows the circular velocities measured at 4~kpc (top) and 8.7~kpc (bottom). The colored points show the circular velocities as indicated by the color bars. The solid white lines show the literature estimates and the dotted white lines show the uncertainties on the literature values. It's reassuring that these regions have a lot of overlap.

We weight the halos so that those with highest weight are consistent with both measurements. Thanks to the smaller uncertainties, the 4~kpc measurement has more constraining power than the 8.7~kpc and most severely restricts the range of halos, but the latter does help to further limit the sample. Mathematically, the weights are calculated via
\begin{equation}
    P_v = G \left( v_\mathrm{circ,halo}, v_\mathrm{circ,obs}, \delta v_\mathrm{circ,obs} \right)
\end{equation}
for each $v_\mathrm{circ}$ measurement.

We can further constrain the halos by considering predictions from cosmological simulations, which demonstrate a clear correlation between their masses and concentrations \citep[e.g.][and many more]{Bullock2001, Dutton2014, Klypin2016, Ragagnin2019}. Although simulations agree that there is a relation between mass and concentration, the exact relation varies from study to study, depending on the simulation used and the cosmology adopted. The parameters used to describe the relation and the mass over which the relation is fitted also varies. We adopt the relation for redshift $z = 0$ from \citet{Dutton2014} as it is provided in virial mass and concentration\footnote{The alternative parameterization is in $M_{200}$ and $c_{200}$ corresponding to the mass-concentration pair where the density enclosed is 200 times the critical density of the universe. It is straightforward to convert between the two \citep{Hu2003} but we prefer to use a relation where this is not necessary.}, and directly covers the mass range in which we are most interested (namely $\sim$ 10$^{10}$ - 10$^{12}$~\Msun). We can use this relation to identify which halos in our space are most likely.

\begin{figure}
    \centering
    \includegraphics[width=\linewidth]{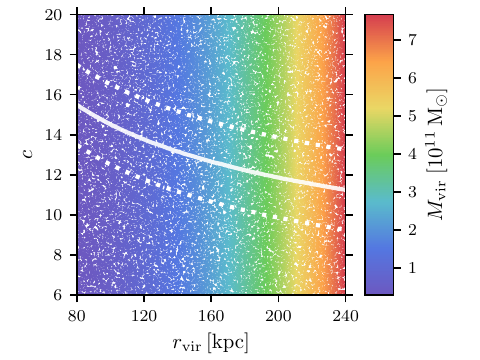}
    \caption{Virial mass $\Mvir$ of the sampled halos. The color indicates the virial mass as indicated by the color bar on the right. Cosmological simulations predict a relation between concentration and virial mass, which we have transformed into concentration and virial radius to plot in this space. The solid white line shows the cosmological relation, and the dotted white lines show our adopted uncertainty on this relation.}
    \label{figure:halos_concmvir}
\end{figure}

\autoref{figure:halos_concmvir} shows the virial masses of the sampled halos, as indicated by the color bar. The solid white line shows the relation from \citet{Dutton2014}, converted from $c-\Mvir$ to $c-\rvir$ to show in the plotted parameter space. We use $H_0 = 69.8 \pm 0.6 \pm 1.6$~km/s/Mpc \citep{Freedman2021}. There will be some uncertainty on this relation due to the uncertainty in $H_0$, and further uncertainties due to the assumptions in the underlying cosmological simulations and analysis, and cosmic scatter with a given simulation, with the latter two considerations being larger than the former. To account for these issues, we adopt a theoretical uncertainty of $\Delta c = 2$ on the relation, with the scatter due to different simulations and within any given simulation guided by Figure~3 of \citet{Ragagnin2019}, albeit over a higher mass range than we use here; this adopted uncertainty is shown as white dotted lines.

We further weight the halos using this relation, that is we give higher weight to the halos whose concentration is closer to the cosmological prediction for a halo of its virial mass. Or mathematically,
\begin{equation}
    P_c = G \left( c_\mathrm{halo}, c_\mathrm{cosmo}, \delta c_\mathrm{cosmo} \right) .
\end{equation}

\begin{figure}
    \centering
    \includegraphics[width=\linewidth]{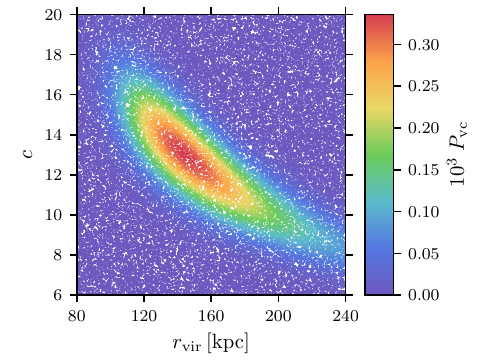}
    \caption{Prior probability of the sampled halos based on their $\vcirc$ and $c-\Mvir$ values. The color indicates the probability as indicated by the color bar on the right.}
    \label{figure:halos_priorprob}
\end{figure}

Now, we wish to combine the weights from the circular velocities and the cosmological simulations. By comparing the white lines in \autoref{figure:halos_vcircs} and \autoref{figure:halos_concmvir}, we can see that these two different constraints intersect, which is useful as it means that there are halos that satisfy both restrictions, but also that they occupy different regions of the parameter space away from their intersection, so together they can constrain the halo sample better than either can alone. Mathematically, we calculate the prior probability of each halo
\begin{equation}
    P_{vc} = P_v (4~\mathrm{kpc}) \times P_v (8.7~\mathrm{kpc}) \times P_c .
\end{equation}
\autoref{figure:halos_priorprob} shows the prior probability $P_{vc}$ for the halo sample, with red indicating the region where the halos have highest weight and purple indicating the regions where the halos have lowest weight.

\subsection{Mass Inside the Most Distant Tracer}
\label{section:massestimate}

Finally, it is time to put all of the pieces together and estimate $\Mrmax$, the mass inside $\rmax$ for our sample.

\begin{figure}
    \centering
    \includegraphics[width=\linewidth]{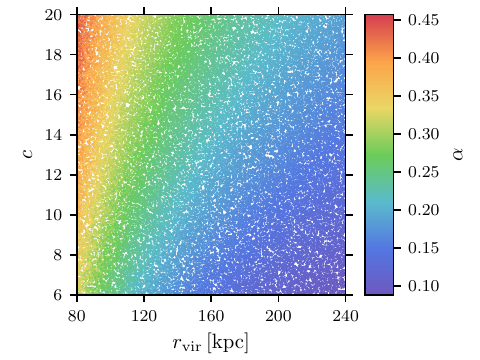}
    \caption{Sample of halos, with points colored by $\alpha$ the slope of the power-law fit to the total (disk plus halo) potential, as indicated by the color bar.}
    \label{figure:halos_alpha}
\end{figure}

At each allowed halo point, we use the fixed $\gamma$ calculated in \autoref{section:tracerdensity}, and draw at random a $\beta$ value from the posterior distribution in \autoref{section:anisotropy}. We also calculate the total (disk plus halo) potential profile and fit for the power-law slope $\alpha$, as described in \autoref{section:potential}. \autoref{figure:halos_alpha} shows the distribution of $\alpha$ for the allowed halos, which span approximately 0.1--0.4.

At each halo point, we also draw a radius $r$ and velocity $v$ for each cluster at random from the samples generated in \autoref{section:dataprocessing}. Finally, we can use the TME in \autoref{eqn:TME} to calculate the mass $\Mtme$ enclosed within $\rmax$ at each halo point. The mass scatter due to the finite number of GCs is implictly included in this approach, but it is also quantified explicitly in \autoref{section:mcsims}. We also show there that the TME does not have an intrinsic bias, not even when applied to a flattened axisymmetric tracer distribution such as that for the LMC GC system.

\begin{figure}
    \centering
    \includegraphics[width=\linewidth]{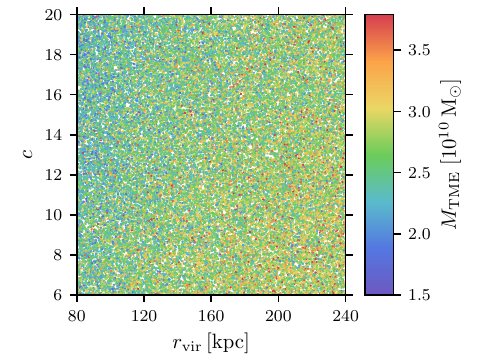}
    \caption{Sample of halos, with points colored by the tracer mass estimate $\Mtme$ of the mass inside $\rmax$.  The $\Mtme$ distribution has a long tail towards higher masses so the upper (red) end of the color bar has been capped at the 99th percentile to better show the variation in $\Mtme$ across the bulk of the sampled halos.}
    \label{figure:halos_tme}
\end{figure}

\autoref{figure:halos_tme} shows the variation in $\Mtme$ calculated by the TME across the halo sample. For visualisation purposes, we have truncated the color bar at the upper end of the mass estimates at the 99th percentile to remove a long, sparsely-populated tail towards higher masses. We can see there is a general trend whereby smaller values of $\alpha$ give a larger mass, but with considerable stochasticity due to the random sampling of $r$ and $v$.

\begin{figure}
    \centering
    \includegraphics[width=0.8\linewidth]{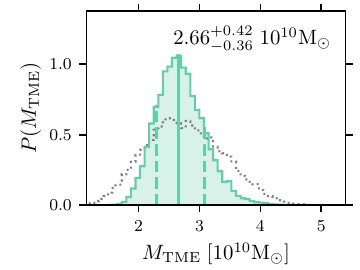}
    \caption{The distribution of estimates for the mass inside $\rmax = \valrmax$~kpc, weighted by the prior halo probabilities. In the top-right corner, we quote the median and 15.9 and 84.1 percentiles as the best estimate and uncertainties. These are shown by the solid and dashed lines respectively. The grey dotted histogram shows the estimated mass within $\valrmax$~kpc from the model grid alone, without the cluster measurements.}
    \label{figure:histogram_tme}
\end{figure}

The distribution of these values gives us an estimate of $\Mrmax$, but we know from \autoref{section:halos} that some halos in the sample are more consistent with our prior knowledge than others. So we weight each $\Mrmax$ estimate by the halo probability $P_{vc}$. \autoref{figure:histogram_tme} shows the resulting weighted distribution of $\Mrmax$ estimates as a histogram. We adopt the median (solid line) and the 15.9 and 84.1 percentiles (dashed lines) as the best estimate and uncertainty, and find
\begin{equation}
    M(<\valrmax \,\mathrm{kpc}) = \valMrmax ^{+\valepMrmax} _{-\valemMrmax} \times 10^{10} M_\odot .
\end{equation}

\begin{figure}
    \centering
    \includegraphics[width=\linewidth]{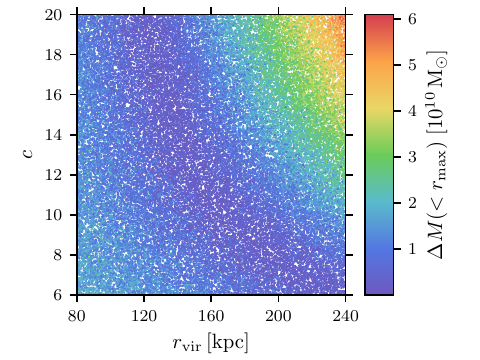}
    \caption{Sample of halos, with points colored by the absolute difference between the mass inside $\rmax$ for the model and the mass inside $\rmax$ estimated from the data, as indicated by the color bar on the right. The stochasticity is due to the random sampling of $r$ and $v$ that is passed to the TME at each halo point.}
    \label{figure:halos_deltam}
\end{figure}

For each halo, we can also estimate the model mass inside $\rmax$, $\Mmodelrmax$. A good consistency check is to verify that there is a region of our sampled parameter space where the enclosed mass of the model halos and enclosed mass estimated from the data are in good agreement. So in \autoref{figure:halos_deltam} we plot the halos with points colored by $\Delta M = | \Mrmax - \Mmodelrmax |$, as indicated by the color bar, where purple indicates regions of good agreement and red indicates regions of poor agreement. Reassuringly there are halos in this space where the $\Delta M$ values are near zero, and, even better, this region intersects with our likely halos as identified by $P_{vc}$.

We can also ask what mass would we estimate inside $\rmax$ if we took the values of $\Mmodelrmax$ across the grid and weighted them by $P_{vc}$ --- that is, if we neglect the cluster measurements and the TME entirely. This distribution is shown as the grey dotted line in \autoref{figure:histogram_tme}. We can see that the mean of the distribution is consistent with our best estimate, but the distribution is $\sim 1.7 \times$ as broad. This highlights the additional constraining power of the cluster measurements and the TME.

\subsection{Virial Mass}
\label{section:virialmass}

We would like to extrapolate the virial mass $\Mvir$ from our TME measurement of $\Mrmax$. Before we do this, however, it is worth revisiting the halo probabilities. We now have more information about each halo.

In \autoref{section:massestimate}, we considered how well each halo $\Mmodelrmax$ matches the measured $\Mrmax$. This gives us another set of weights for each halos.

$\Delta M$, as shown in \autoref{figure:halos_deltam}, provides a useful visualisation to look at the overall trends, but there is still considerable stochasticity amongst near neighbors due to the random sampling of $r$ and $v$, as we saw in \autoref{figure:halos_tme}. So instead we consider how well each halo $\Mmodelrmax$ agrees with the weighted distribution of $\Mrmax$ values shown in \autoref{figure:histogram_tme}, which has the scatter in $r$ and $v$ convolved in and also accounts for the prior probabilities of the halos. The distribution is mildly asymmetric so, similarly to \autoref{section:anisotropy}, we fit a function that is the sum of $N$ Gaussians to the distribution of $\Mrmax$ estimates, where $N=2$ this time. Then the probability of any given halo is
\begin{equation}
    \Pmass = \sum^2_k f_k G \left( m_k - M_\mathrm{model}, w_k \right) .
    \label{equation:likelihoodmass}
\end{equation}
where $M_\mathrm{model}$ is $\Mmodelrmax$ for the halo and $f_k$, $m_k$, and $w_k$ are the parameters of the Gaussian function as defined in \autoref{equation:gaussians}.

We can combine the probabilities from the enclosed mass $\Pmass$ with our prior probabilities from the circular velocities and cosmological simulations $\Pvc$ to get a posterior probability
\begin{equation}
    \Phalo = \Pvc \times \Pmass .
\end{equation}

\begin{figure}
    \centering
    \includegraphics[width=\linewidth]{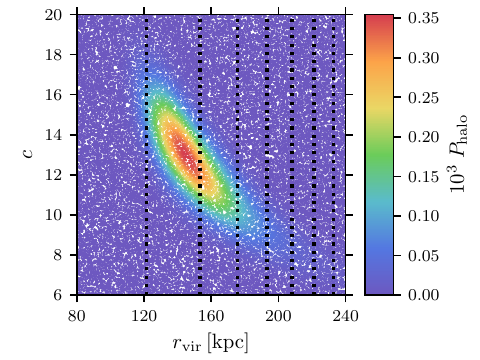}
\caption{Sample of halos, with points colored by the probability of each halo, based on the enclosed mass within $\rmax$, the observed circular velocity, and consistency with cosmological predictions. Vertical dotted lines mark virial masses $\Mvir$ of $1-7 \times 10^{11}$~\Msun.}
    \label{figure:Phalo}
\end{figure}

\autoref{figure:Phalo} shows the distribution of $\Phalo$ values across the halo sample from high in red to low in blue. The vertical dotted lines mark virial masses of (1,2,3,4,5,6,7) $\times 10^{11}$~Msun. We see that this distribution is tighter than the distribution of $\Pvc$ in \autoref{figure:halos_priorprob}, in particular, low and high $\rvir$ and, thus, low and high $\Mvir$ halos are disfavored.

\begin{figure}
    \centering
    \includegraphics[width=0.8\linewidth]{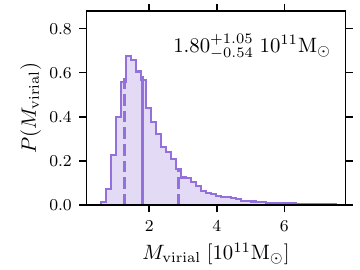}
    \caption{The weighted distribution of virial mass estimates extrapolated from the mass inside $\rmax$ measured from the data. In the top-right corner, we quote the median and 15.9 and 84.1 percentiles as the best estimate and uncertainties. These are shown by the solid and dashed lines respectively.}
    \label{figure:Mvirial}
\end{figure}

The final step is to consider the distribution of virial masses $\Mvir$ across the sample, with each point weighted by $\Phalo$. \autoref{figure:Mvirial} shows the weighted distribution of virial mass estimates for the sample. We adopt the median (solid line) and 15.9 and 84.1 percentiles (dashed lines) of the weighted distribution as the best estimate and uncertainty, and find
\begin{equation}
    \Mvir = \valMvir ^{+\valepMvir} _{-\valemMvir} \times 10^{11} M_\odot .
\end{equation}

\subsection{Halo Properties}

Using a similar approach as that used to estimate the virial mass, we can also offer a best estimate of the virial radius $\rvir$ and concentration $c$ of the LMC's DM halo. That is, we ask what are the properties of halos for which the models best match the observations, and we do this by considering the weighted distribution of $\rvir$ and $c$ across the sample using the $\Phalo$ values as weights.

\begin{figure}
    \centering
    \includegraphics[width=0.8\linewidth]{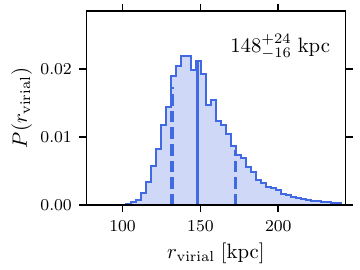}
    \includegraphics[width=0.8\linewidth]{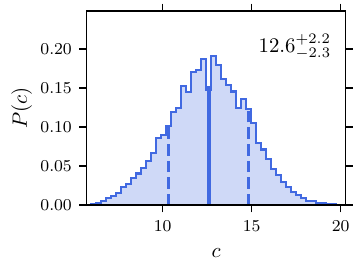}
    \caption{The distribution of virial radii (top) and concentrations (bottom) weighted by the probability of each halo. In the top-right corner of each panel, we quote the median and 15.9 and 84.1 percentiles of the distribution as the best estimate and uncertainties. These are shown by the solid and dashed lines respectively.}
    \label{figure:haloprops}
\end{figure}

\autoref{figure:haloprops} shows the weighted distributions of the $\rvir$ (top) and $c$ (bottom) values. We adopt the median (solid lines) and 15.9 and 84.1 percentiles (dashed lines) of the weighted distribution as the best estimate and uncertainty, and find virial radius
\begin{equation}
    \rvir = \valrvir ^{+\valeprvir} _{-\valemrvir} \; \mathrm{kpc} ,
\end{equation}
and concentration
\begin{equation}
    c = \valc ^{+\valepc} _{-\valemc} .
\end{equation}

\begin{figure}
    \centering
    \includegraphics[width=0.8\linewidth]{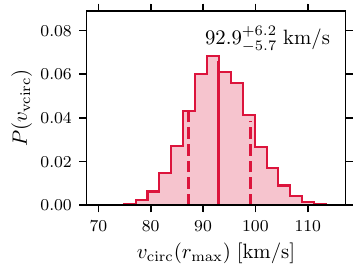}
    \caption{The distribution of circular velocities at $\rmax$ weighted by the probability of each halo. In the top-right corner, we quote the median and 15.9 and 84.1 percentiles of the distribution as the best estimate and uncertainties. These are shown by the solid and dashed lines respectively.}
    \label{figure:Vrmax}
\end{figure}

We can also estimate the circular velocity $\vcirc$ at $\rmax$ using the same method. The weighted distribution of circular velocity estimates is shown in \autoref{figure:Vrmax}. Once again, we adopt the median (solid lines) and 15.9 and 84.1 percentiles (dashed lines) of the weighted distribution as the best estimate and uncertainty, and find a circular velocity at $\rmax$
\begin{equation}
    \vcirc (\valrmax) = \valvrmax ^{+\valepvrmax} _{-\valemvrmax} \; \mathrm{km/s} .
\end{equation}

\section{Discussion}
\label{section:discussion}

Now that we have made estimates of the mass and other halo properties from the observations, let us consider how they compare to previous estimates, and the implications they have for our understanding of the LMC and the MW.

\subsection{Mass from Tracers}
\label{section:discussrmax}

We begin with the mass we can estimate most directly from the observations. We have estimated the mass inside the radius of our furthest tracer to be $M(<\valrmax \,\mathrm{kpc}) = \valMrmax ^{+\valepMrmax} _{-\valemMrmax} \times 10^{10}$~\Msun.

No other study has made a mass estimate at this radius before, although many have estimated the mass in the vicinity. To compare our estimate with others, we need to make some reasonable assumptions about how the mass changes over small radial ranges in this region. Typically, we assume a nearly isothermal sphere, in which case mass would increase linearly. This assumption is good enough for our purposes here of comparing our estimate against previous studies, but is likely more reliable over smaller radial ranges than larger ones.

\citet{Schommer1992} used LOSVs of LMC GCs to estimate a mass of $1.5-2.5 \times 10^{10}$~\Msun within 8$^\circ$. At our adopted distance of the LMC (see \autoref{table:lmcprops}), this corresponds to $\sim$7~kpc. Extrapolating linearly, this would imply $M (<13.2 \, \mathrm{kpc}) = 2.8-4.7 \times 10^{10}$~\Msun, a little on the high side of our estimate, but not inconsistent. This offset may be because we used full 3-dimensional velocity information (PMs and LOSVs) for our estimate, while \citet{Schommer1992} had only 1-dimensional LOSV information.

\citet{Kim1998} studied HI in the LMC to estimate a mass of $\sim0.35 \times 10^{10}$~\Msun within 4~kpc. Extrapolating linearly, this would imply $M (<13.2 \, \mathrm{kpc}) \sim 1.2 \times 10^{10}$~\Msun, a little on the low side of our estimate and only marginally consistent. This may indicate that the HI gas does not paint a full picture of the LMC's inner regions.

More recently, \citet{vdMarel2014} found $M (<8.7 \, \mathrm{kpc}) = 1.7 \pm 0.7 \times 10^{10}$~\Msun. Again, assuming that mass increases linearly from 8.7~kpc to $\valrmax$~kpc, this would predict a mass of $M (<13.2 \, \mathrm{kpc}) = 2.54 \pm 1.05 \times 10^{10}$~\Msun, in very good agreement (well within 1-sigma) of the value we find.

Overall, our estimate lies within the range spanned by previous estimates, and is in very good agreement with the value that required least extrapolation and was the most complete data set of the previous studies.

\subsection{Circular Velocity}
\label{section:discussvrmax}

It is also useful to consider the circular velocity at $\rmax$, which we have measured to be $\vcirc$($\valrmax$~kpc) = $\valvrmax ^{+\valepvrmax} _{-\valemvrmax}$~km/s.

Recall, we used the circular velocities at 4~kpc and 8.7~kpc in \autoref{section:halos} to provide constraints on the allowed halos. These values were $v_\mathrm{circ}$(4~kpc) = 85 $\pm$ 5~km/s \citep[calculated from][]{GaiaCollaboration2021} and $\vcirc$(8.7~kpc) = 91.7 $\pm$ 18.8~\kms \citep{vdMarel2014}.

These values imply a rotation curve that is mostly flat and, possibly, slightly rising over the range of the GC tracers.

\subsection{Total Mass of the LMC}
\label{section:discusstotal}

Now we turn to the total mass of the LMC. We have estimated a value $\Mvir = \valMvir ^{+\valepMvir} _{-\valemMvir} \times 10^{11}$~\Msun. A number of previous studies have estimated the LMC mass, using a wide range of observations and techniques. How does our estimate compare?

The orbit of the LMC within the MW and within the LG can help constrain its mass. In a PM study of the LMC, \citet{Kallivayalil2013} explored a range of MW and LMC masses, with fixed SMC mass, and found that large LMC masses ($\Mlmc > 1 \times 10^{11}$~\Msun) were preferred in their models for the LMC and SMC to have been a bound pair\footnote{At the time of the study, the MW mass was also highly uncertain, and they favored a value lower than $1.5 \times 10^{12}$~\Msun; later works have reduced the uncertainty on the MW mass to be consistent with this prediction \citep[e.g.][]{Watkins2019}, strengthening the LMC-SMC prediction as well.}. This is consistent with what we find, with only $\sim$4\% of the (weighted) halos in \autoref{figure:Mvirial} being below $1 \times 10^{11}$~\Msun.

Later, \citet{Penarrubia2016} applied the timing argument to the MW-M31-LMC system and estimated the LMC's mass to be $\Mlmc = 2.5 _{-0.8} ^{+0.9} \times 10^{11}$~\Msun, in very good agreement (well within 1-$\sigma$) of what we find here.

Studying the effect of the LMC on the MW can also help constrain its mass. \citet{Laporte2018} studied the warp in the MW disk induced by the LMC, and favored a model where the LMC mass is $2.5 \times 10^{11}$~\Msun. More recently, \citet{Erkal2021} studied the response of the MW halo to the infall of the MW and favor an LMC mass of $M_\mathrm{LMC} = 1.5 \times 10^{11}$~\Msun. The former is in very good agreement with our study (consistent within 1-$\sigma$), the latter is less so, but still agrees within 2-$\sigma$.

Stellar streams in the MW halo also have a lot of constraining power. \citet{Erkal2019} used the path of the Orphan Stream to estimate the total mass of the LMC at $\Mlmc = 1.38 ^{+0.27} _{-0.24} \times 10^{11}$~\Msun. Similarly, \citet{Vasiliev2021} used the path of the Sagittarius Stream to estimate the mass of the LMC to be $\Mlmc = 1.3 \pm 0.3 \times 10^{11}$~\Msun. More recently, \citet{Shipp2021} offered two LMC mass estimates. Using an ensemble of streams they found a mass ranging from $\Mlmc \sim 1.4-1.9 \times 10^{11}$~\Msun, and using their most-constraining stream they found $\Mlmc = 1.88 ^{+0.35} _{-0.40} \times 10^{11}$~\Msun. All of these estimates are a little low compared with our estimate, but are still within or just outside of 1-$\sigma$ consistent.

So far, these estimates have all be focused directly on the LMC, but there is one further cosmological argument. The stellar mass of the LMC is $\sim 2.5 \times 10^9$~\Msun \citep{Kim1998}. Abundance-matching studies \citep[e.g.][]{Behroozi2013} suggest that a galaxies of this stellar mass typically inhabit halos of $\sim 2-3 \times 10^{11}$~\Msun, which is also in good agreement with what we find here.

\begin{figure*}
    \centering
    \includegraphics[width=\linewidth]{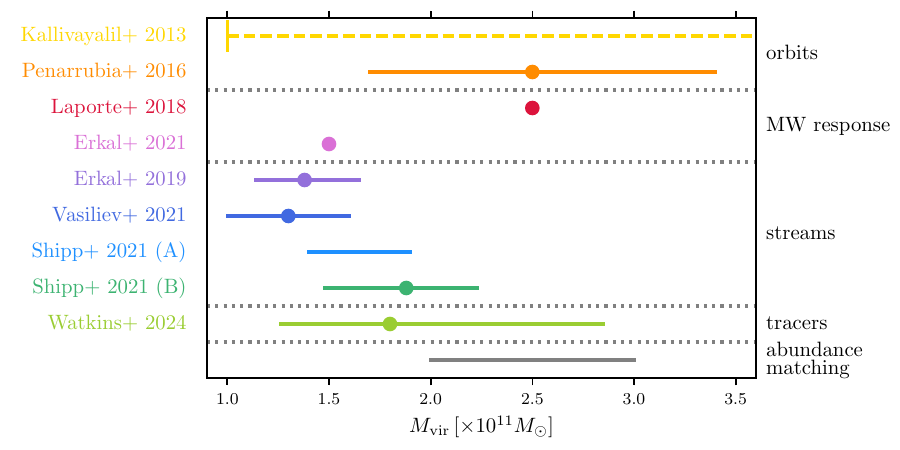}
    \caption{Total mass estimates of the LMC from previous studies and this work. Estimates are given as lower limits, single values, values with uncertainties, or ranges, depending on the study. Sources are provided on the left, the methods used are shown on the right. Horizontal grey dotted lines separate different methods.}
    \label{figure:literature}
\end{figure*}

A summary of all these estimates is presented in \autoref{figure:literature}. It is reassuring that all of these studies--the present study included--have used widely different methods or observations but are producing consistent estimates. It is also worth noting that our method, which extrapolates the virial mass from data much further in, is generally in good agreement with methods that estimate the total mass directly, albeit with larger uncertainties.

Assuming a MW mass of $1.1 \pm 0.2 \times 10^{12}$~\Msun \citep[a value summarising many recent estimates from][]{Sawala2023}, our LMC virial mass estimate of $\valMvir ^{+\valepMvir} _{-\valemMvir} \times 10^{11}$~\Msun\ means that the LMC is $17 ^{+10} _{-6}$\% of the mass of the MW. Further assuming an M31 mass of $1.5 \pm 0.5 \times 10^{12}$~\Msun \citep[again a summary value from][]{Sawala2023}, and an approximate M33 mass of $4.3 \pm 0.1 \times 10^{11}$~\Msun \citep{Corbelli2014}, the LMC contributes $6 ^{+3} _{-2}$\% of the mass in the LG. This reinforces the point that the LMC is a non-negligible contributor to both the MW and LG potentials.

\section{Conclusions}
\label{section:conclusions}

We have used a kinematic catalog of 30 GCs in the LMC to estimate the anisotropy and mass of the LMC within $\rmax = \valrmax$~kpc. We have also used this information to extrapolate the virial mass of the LMC.

We find that the anisotropy of the GC system defined in a spherical coordinate system centred on the LMC is $\beta = \valbeta ^{+\valepbeta} _{-\valembeta}$ within $\valrmax$~kpc. This negative bias to the anisotropy indicates that the tangential velocities are larger than the radial velocities, which is consistent with the GC system  being flattened, rotating, and pressure supported.

Applying a tracer mass estimator to the GC sample, we find a mass $\Mrmax = \valMrmax ^{+\valepMrmax} _{-\valemMrmax} \times 10^{10}$~\Msun, consistent with previous studies that have estimated masses at similar radii. Using a grid of NFW halos to extrapolate out to the virial radius, this implies that the LMC has a virial mass $\Mvir = \valMvir ^{+\valepMvir} _{-\valemMvir} \times 10^{11}$~\Msun, also consistent with previous studies estimating the LMC's total mass that have used a variety of different methods.

Until recently, lack of PMs has been a severely limiting factor in our ability to do these kinds of studies, but now --- thanks to both \textit{HST} and \textit{Gaia} --- we are in an era where PMs are often more readily available than LOSVs or distances. This is the case here, where we were unable to include a number of LMC GCs in our analysis for which PMs had been measured because they lacked either distance or LOSV information, or both. Even for the GCs we did include, the distance and LOSV estimates often had large uncertainties. Acquiring more and more accurate distance and LOSV information of the LMC GCs, will increase the accuracy with which we can estimate the mass, both due to the smaller uncertainties and the increased sample size.

With that said, we have made a number of assumptions here, that were good enough to get us to an answer and to shed new light on the LMC, but that are not strictly true. Further improvements could be made by performing more sophisticated modelling that can account for greater complexity than we have been able to here. However, our final uncertainties on the LMC virial mass are dominated by the shot noise associated with having a finite sample of clusters, and the systematic uncertainties associated with making a large extrapolation our to the virial radius. Hence, more sophisticated models are not likely to lead to a better constrained mass.

\begin{acknowledgments}

We thank the anonymous referee for their helpful suggestions. LLW wishes to thank Gurtina Besla, Ekta Patel, Mark Fardal, and Eduardo Vitral for very useful conversations related to this work.

Support for this work was provided by a grant for \textit{HST} program GO-15633 provided by the Space Telescope Science Institute, which is operated by AURA, Inc., under NASA contract NAS 5-26555.

This research made use of Astropy\footnote{\url{http://www.astropy.org}}, a community-developed core Python package for Astronomy. This research has made use of NASA's Astrophysics Data System Bibliographic Services.

This project is part of the HSTPROMO (High-resolution Space Telescope PROper MOtion) Collaboration\footnote{\url{http://www.stsci.edu/~marel/hstpromo.html}}, a set of projects aimed at improving our dynamical understanding of stars, clusters and galaxies in the nearby Universe through measurement and interpretation of proper motions from \textit{HST}, \textit{Gaia}, \textit{JWST}, and other space observatories. We thank the collaboration members for the sharing of their ideas and software.

\end{acknowledgments}

\vspace{5mm}

\software{
	Astropy \citep{astropy2013, astropy2018},
	\textsc{emcee} \citep{ForemanMackey2013},
	IPython \citep{ipython2007},
    galpy \citep{Bovy2015},
	Matplotlib \citep{matplotlib2007},
	NumPy \citep{numpy2020},
    \textsc{scalefree} \citep{deBruijne1996},
	SciPy \citep{scipy2020}
}

\appendix

\section{Bias and Scatter in the Tracer Mass Estimator}
\label{section:mcsims}

In general, a tracer mass estimator will have both a bias and a scatter when estimating the true mass of a system. We use Monte-Carlo simulations to quantify this, both for the case of a spherical distribution of tracer particles and for an axisymmetric distribution.

\subsection{Spherical Distributions}
\label{section:spherical}

Our simulations for a spherical distribution of tracer particles are described in full in \citet{Watkins2010}, but we briefly summarise them here. We draw $N_S$ sets of $N_T$ tracers in $\rmin$ to $\rmax$ from a power-law density distribution with slope $\gamma$. We draw their velocities from the distribution function given in \citet{Evans1997}, assuming a power-law potential with slope $\alpha$ and an anisotropy $\beta$.

For these simulations, we use $\rmin = \valrmin$~kpc and $\rmax = \valrmax$~kpc to match the data. We use the fixed $\gamma = \valgamma$ we estimated in \autoref{section:tracerdensity}, the median $\beta = \valbeta$ we estimated in \autoref{section:anisotropy}, and $\alpha = \valalpha$, the median of the distribution generated in \autoref{section:potential}. The results of the simulations do not vary strongly with the particular choice of $\alpha$, $\beta$ and $\gamma$; so these representative values are good enough to cover the whole halo sample. Finally, $N_T = 30$, the number of tracers in our sample, and we chose to use $N_S = 10,000$ to give robust statistics.

\begin{figure}
    \centering
    \includegraphics[width=0.8\linewidth]{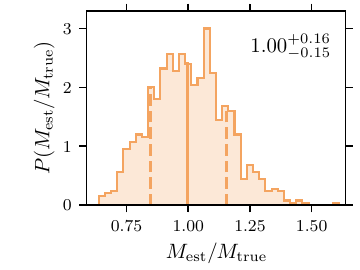}
    \caption{Results from a Monte Carlo simulation of 10,000 sets of 30 tracers. The histogram shows the ratio of the mass inside $\rmax$ estimated using the TME to the true mass of the simulation. That this ratio, on average, is close to 1 indicates there is little bias in the estimator, and the spread in values indicates the uncertainty in the estimator.}
    \label{figure:mcsims}
\end{figure}

For each set of $N_T$ tracers, we use the TME to estimate $\Mtme$. As we also know the true mass of the model $\Mtrue$, we can calculate the ratio $f = \Mtme / \Mtrue$. The resulting distribution of $f$ values in shown in \autoref{figure:mcsims}. We use the median (shown as the solid line), and the 15.9 and 84.1 percentiles (shown as dashed lines) to represent the distribution and find $f = \valf ^{+\valepf} _{-\valemf}$. The median of unity indicates that the there is very little to no bias in the estimator and that it is able to recover the true mass very well on average. The scatter in $f$ gives us an indication of the contribution the TME makes to the uncertainty in our estimate for $\Mrmax$. This scatter is included in our analysis of \autoref{section:massestimate}, since we include random drawing of clusters samples for all halo points. 

For reference, in the limiting case of a spherical isotropic ($\beta = 0$) $\rho \propto r^{-3}$ distribution in a logarithmic potential ($\alpha = 0$), the ratio $f$ can be expressed analytically as $f = \overline{v^2} / V_0^2$. Here $V_0$ is the circular velocity of the potential, and the isotropic velocity dispersion in each direction equals $\sigma = V_0 / \sqrt{3}$. This yields a mean $\overline{f} = 1$ and RMS scatter $\Delta f = \sqrt{2/3N_T}$. For $N_T = 30$ this implies $\Delta f = 0.15$, in good agreement with our simulations.

\subsection{Flattened Distributions}
\label{section:flattened}

\citet{Bennet2022} showed that the LMC globular cluster system has a flattened configuration that rotates like the stellar disk. The median $|z|$ for the GCs in their Table~3 is 1.27~kpc versus a median $R$ of 3.32~kpc, implying an axial ratio of $q \approx 0.4$. Hence it is important to also assess how the axial ratio of the tracer distribution affects the accuracy of the TME.

To address this question we use the \textsc{scalefree}\footnote{\url{https://gitlab.com/eduardo-vitral/scalefree}} models of the \citet{deBruijne1996}. They studied dynamical models with varying velocity dispersion anisotropy for a power-law tracer distribution of axial ratio $q$ embedded in a spherical power-law gravitational potential. This situation is similar to that studied by \citet{Watkins2010}, but now with an additional flattening of the tracer distribution. There is a wide range of three-integral phase-space distribution functions (DFs) that is consistent with such a configuration. However, some special cases can be derived analytically, especially for gravitational potentials that are either logarithmic ($\alpha=0$) or Keplerian ($\alpha=1$). The case~I DFs of \citet{deBruijne1996} have spatially-variable anisotropy. They are anisotropic generalizations of flattened $f(E,L_z)$ models (where $E$ is the energy of an orbit and $L_z$ its angular momentum), which they include as a special case. The case~II DFs instead correspond to models with constant anisotropy $\beta$ (as defined by \autoref{eqn:anisotropy}) throughout the system (but with the ratio $\overline{v_{\theta}^2} / \overline{v_{\phi}^2}$ varying as function of angle $\theta$ is the meridional $(R,z)$ plane). All the models have a free parameter $\beta_p$ that governs the amount of velocity anisotropy. For the case~II DFs $\beta = \beta_p$ everywhere, whereas for the case~I DFs this is true only on the symmetry axis.

To assess the accuracy of the TME we again Monte-Carlo draw samples of $N_T$ tracer particles from a density distribution of power-law slope $\gamma$, but now with axial ratio $q$. We draw a velocity for each particle from the velocity ellipsoid for a chosen DF (either case~I or~II, and for a given choice of $\beta_p$ and $\alpha$). We then infer the ratio $f$ as before and repeat this in Monte-Carlo fashion to derive both the bias and scatter inherent in using the TME for a flattened tracer distribution.

\begin{figure}
    \centering
    \includegraphics[width=\linewidth]{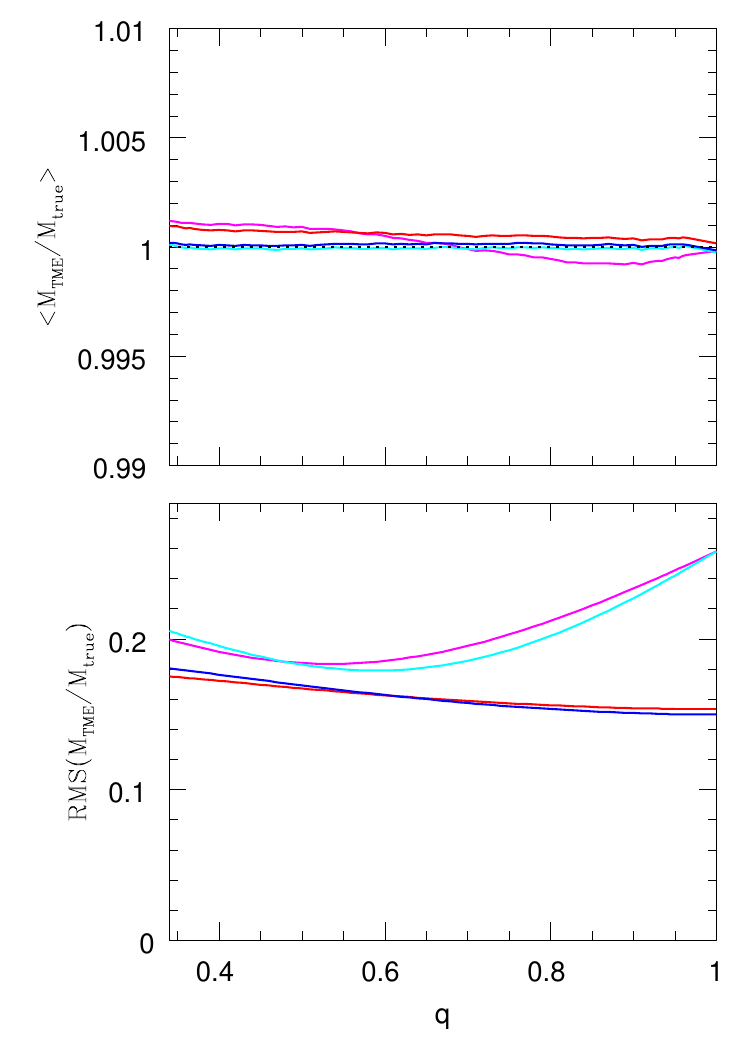}
    \caption{Bias (top) and scatter (bottom) in the TME when applied to an axisymmetric tracer density of axial ratio $q$. Curves are for either a spherical logarithmic potential ($\alpha = 0$; blue and cyan) or a spherical Kepler potential ($\alpha=1$; red and  magenta). The DF is either case~I with variable anisotropy (magenta and cyan) or case~II with fixed anisotropy (red and blue). The flattening of the density does not introduce a significant bias, and does not significantly change the scatter.}
    \label{figure:TMEaxisym}
\end{figure}

\autoref{figure:TMEaxisym} shows the results as a function of $q$ for case~I and case~II DFs with either $\alpha =$ 0 or 1. These four curves bracket the best-fit potential (centered around $\alpha = 0.21$) with either constant or spatially-varying anisotropy. In the case~II DFs with constant anisotropy we used $\beta_p = -0.72$, equal to the observed value derived in \autoref{section:anisotropy}. In the case~I DFs instead we used $\beta_p = 1$. These models are radially anisotropic on the symmetry axis, but tangentially anisotropic towards the equatorial plane. When $q=0.4$, the 3D-integrated anisotropy for the latter models is $\beta = -0.99$ for a logarithmic potential and $\beta = -1.49$ for a Keplerian potential. These values are consistent with the observed value at $1\sigma$ confidence.

The upper panel of \autoref{figure:TMEaxisym} shows that for the case of a logarithmic potential, the TME has no bias even when applied to a highly-flattened system. In a Kepler potential there is a small bias, but it is $\lesssim 0.1$\% at all values of the axial ratio, and for both DF cases. This is well below the random uncertainty in any of our estimates. So while this analysis falls well short of a full exploration of all possible axisymmetric DFs, it does suggest that the fact that the LMC GC system is flattened does not introduce significant biases in our mass estimates.

The lower panel of \autoref{figure:TMEaxisym} shows that the flattening of the tracer density does introduce some added scatter in the TME. At $q=0.4$ the scatter for the four models ranges from $\Delta f = 0.18$--$0.20$, slightly larger than the value $\Delta f = 0.15$ derived for a spherical model in \autoref{section:mcsims}. However, this difference is well below the random uncertainty in any of our estimates. So this added scatter also does not meaningfully affect our mass estimates derived in the main text.

\bibliography{References}

\end{document}